\documentclass[10pt,conference]{IEEEtran}

\usepackage{amsmath}
\usepackage{subfigure}
\usepackage{graphicx}
\usepackage{latexsym}
\usepackage{amssymb,amsbsy,verbatim,array}
\usepackage{epsfig}
\usepackage{cite}
\usepackage{color}
\usepackage{algorithm}
\usepackage{algorithmicx}
\usepackage{algpseudocode}

\begin{document}

\title{STAC: Simultaneous Transmitting and Air Computing in Wireless Data Center Networks}
\author{
\authorblockN{Shengli Zhang}
\authorblockA{Faculty of Information Engineering\\
Shenzhen University\\
Shenzhen, China\\
Email: zsl@szu.edu.cn} \and
\authorblockN{Xiugang Wu}
\authorblockA{Department of Electrical Engineering\\
Stanford University\\
CA, US\\
Email: x23wu@standord.edu}\and
\authorblockN{Ayfer Ozgur}
\authorblockA{Department of Electrical Engineering\\
Stanford University\\
CA, US\\
Email: aozgur@standord.edu}
}

\maketitle
\begin{abstract}

The data center network (DCN), wired or wireless, features large amounts of Many-to-One (M2O) sessions. Each M2O session is currently operated based on Point-to-Point (P2P) communications and Store-and-Forward (SAF) relays, and is generally followed by certain further computation at the destination.
%typically a weighted summation of the received digits.
Different from this separate P2P/SAF-based-transmission and computation strategy, this paper proposes STAC, a novel physical layer scheme that achieves Simultaneous Transmission and Air Computation in wireless DCNs. In particular, STAC takes advantage of the superposition nature of electromagnetic (EM) waves, and allows multiple transmitters to transmit in the same time slot with appropriately chosen parameters, such that the received superimposed signal can be directly transformed to the needed summation at the receiver. Exploiting the static channel environment and compact space in DCN, we propose an enhanced Software Defined Network (SDN) architecture to enable STAC, where wired connections are established to provide the wireless transceivers external reference signals. Theoretical analysis and simulation show that with STAC used, both the bandwidth and energy efficiencies can be  improved severalfold.

%
%Flexible and high bandwidth wireless links are preferred for
%the random traffic in data center networks. Observing typical
%jobs that data from several nodes are transmitted to
%one receiver to obtain a linear combination, we propose to
%make the sources transmitting at the same time such that
%the superimposed signal at the receiver comes into being the
%needed form, referred to as STAC. As a result, the transmission
%and computation are completed in one shot. To enable
%STAC, we propose to use an centralized wired control network
%to provide control information and external reference
%signals to guarantee the configuration and synchronization
%of the wireless transceivers, taking the advantages of static
%channel conditions and compact space in data centers. Applying
%the STAC mechanism to data center networks, the
%routing and scheduling problems, facing totally new challenges,
%are also discussed in this paper. Simulation shows
%that both energy efficiency and bandwidth efficiency can be
%improved by several folders.
\end{abstract}

%\category{C.2.1}{Network Architecture and Design}{Network communications, Wireless communications}
%A category including the fourth, optional field follows...
%\category{D.2.8}{Software Engineering}{Metrics}[complexity measures, performance measures]

%\terms{Theory, Design}

%\keywords{big data center, wireless, network coding} % NOT required for Proceedings

\section{Introduction}

A modern Data Center (DC) typically consists of a large dedicated cluster of commercial computers (work nodes) that are housed together to store/process big files in a parallel manner. The characteristic of parallel storing/processing requires frequent communications among the work nodes, which are accomplished through Data Center Networks (DCNs). Today, DCN is the principle bottleneck in large DCs \cite{Al-Fares_DCN_architecture_2008}.
%
%With the rapidly developing of big data processing and cloud computing, distributed data center (DC) is becoming an efficient and popular form, where thousands of cheap computers are put together to store and process the big data. In particular, the storage and processing jobs are achieved by each work node in a parallel way and information is exchanged (communication) among the work nodes when necessary. Observing the Many-to-One (M2O) transmission and the linear computation in DC, we propose STAC, a new transmission mechanism to integrate separated communications and computations into one shot to significantly improve the performance.
Despite of its maturity in deployment and high bandwidth, the wired DCN has a few critical problems such as flexibility, cabling complexity, device cost, over subscription, etc. These problems highly limit the efficiency and scalability of the DCN and are being exacerbated provided that a huge amount of information needs to be stored/processed within the DC and exchanged through the DCN in today's big data age.

To address this issue, some works \cite{conf:hotnets:KandulaPB09, ramachandran:2008, Zhou:2012:MMC:2342356.2342440, Hamedazimi:2014:FRW:2619239.2626328, Halperin:2011:ADC:2018436.2018442, Shin:2013:FCW:2578911.2578935} studied the possibility of constructing wireless DCNs using high frequency electromagnetic (EM) waves.
The 60 GHz techniques were {suggested for realizing} wireless DCN links with bandwidth comparable to wireline connections \cite{conf:hotnets:KandulaPB09, ramachandran:2008}, while the blockage and directivity problems associated with the EM waves can be significantly mitigated by utilizing the strategies of ceiling reflection and 3D beamforming  \cite{Zhou:2012:MMC:2342356.2342440}.
%(cf. Section 3.1)
Free-space optical DCN communications were also investigated, and were shown to achieve some further improvements, including higher bandwidth and nearly perfect directivity \cite{Hamedazimi:2014:FRW:2619239.2626328}. On the other hand, from the structural perspective, the work \cite{Halperin:2011:ADC:2018436.2018442} considered augmenting the wired DCN with added wireless flyways, and \cite{Shin:2013:FCW:2578911.2578935} demonstrated that a completely wireless DCN with a Cayley structure is feasible and performs even better than the wired DCN.

\begin{figure}[tbp]
\begin{center}
\includegraphics[width=3.5in]{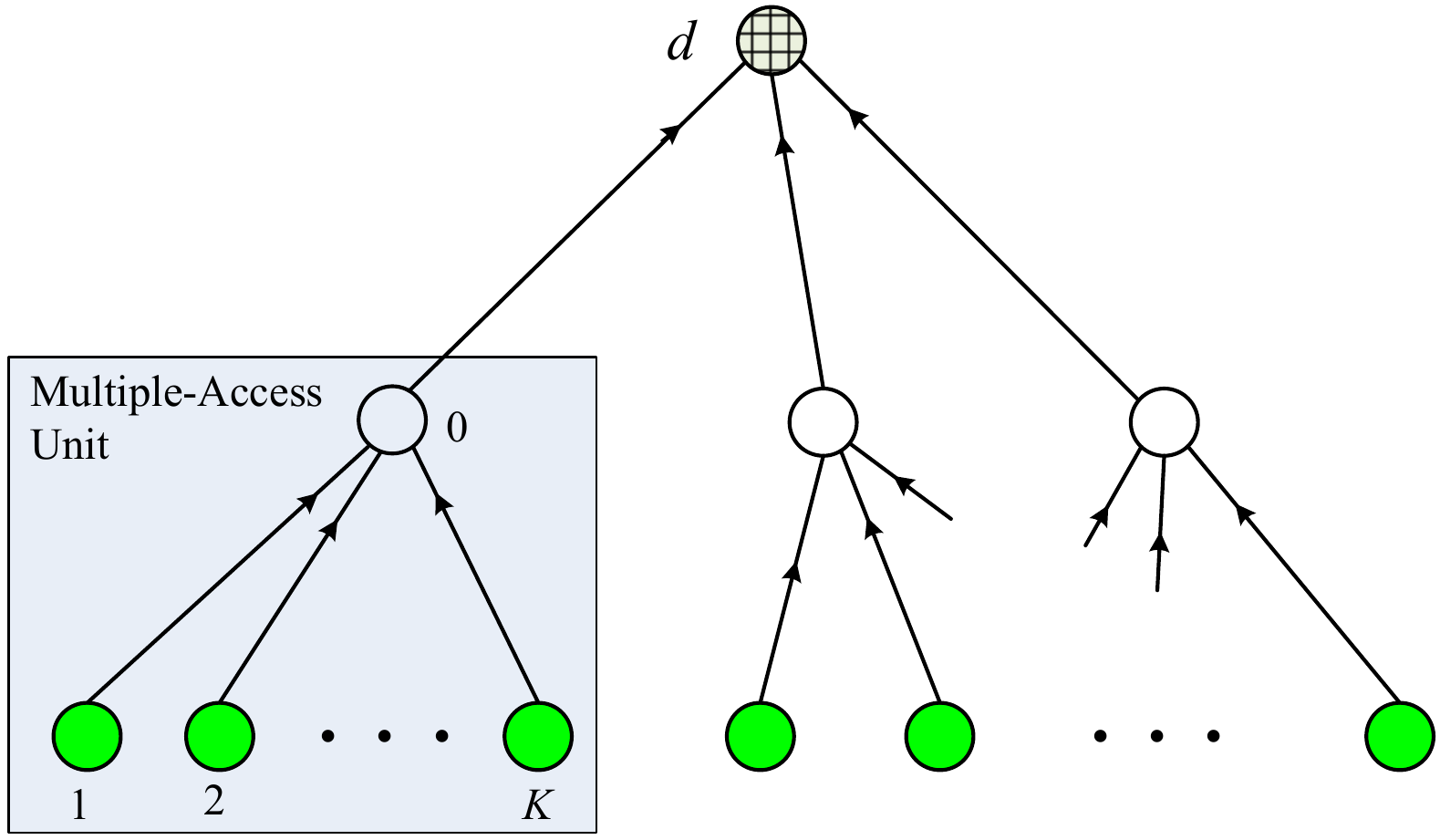}
\caption{Illustration of one M2O session. Source nodes (solid circles) transmit information to destination $d$ via relay nodes (hollow circles).}
\label{fig_illustrating_model}
\end{center}
\end{figure}

%\subsection{Challenging P2P and SAF Strategies}
%{\color{red}This paper aims to further the research on the wireless DCN by challenging two of its fundamental transmission strategies, namely, the Point-to-Point (P2P) communication and the Store-and-Forward (SAF) relay strategies. These two strategies are widely adopted in today's ubiquitous wireless networks, e.g., in the cellular and 802.11 networks, but here we try to argue that they may not be suitable for DCNs due to some distinguishing features of the traffic and computation occurring within  DCs.}

Wireless DCN is different from today's ubiquitous wireless networks, through traffic patterns to network structures. These differences can provide new challenges, as well as possibilities, to design more efficient wireless DCNs.
One particular challenge in DC is the large amounts of Many-to-One (M2O) sessions, which brings some new problems for DCNs, especially with the Point-to-Point (P2P) communication and the Store-and-Forward (SAF) relay strategies. The M2O sessions arise from various DC applications, e.g., Google File System (GFS) \cite{Ghemawat_GFS_2003} and MapReduce \cite{Dean_Mapreduce_2008} framework. Due to the limited transmission range of high frequency EM waves, these M2O sessions are operated through multi-hopping over hierarchical multiple-access units as shown in Fig. \ref{fig_illustrating_model}, where each hop is based on the P2P communication and followed by the SAF relay to the next hop. Specifically, in the multiple-access unit as depicted, with Time Division Multiple-Access (TDMA), the source nodes $1,2,\ldots, K$ successively transmit their information digits to the relay node $0$ in  different time slots with P2P strategy, and the relay stores all its received digits in the buffer before  forwarding them to the destination $d$. Since the node 0's buffer and input/output bandwidth are shared by all the $K$ source nodes, the transmission performance could be poor, especially when $K$ is large. The nearer to the destination, the severer this problem will be, as the information that needs to be transmitted accumulates along the way. In fact, the problem of TCP throughput collapse caused by M2O transmissions in data center networks have been noted as incast problem \cite{Nagle_incast_2004}.

\subsection{A New Scheme: STAC}

%Given these challenges to the current DCN transmission strategies, how do we make improvements? Two natural changes that one may come up with, corresponding to improving P2P and SAF, respectively, are the following.
%\begin{enumerate}
%  \item The sources transmit their digits in the \emph{same} time slot, and the receiver recovers the digits based on \emph{multi-user detection}.
%  \item The relay performs \emph{computation} (weighted summation typically) of the recovered digits, and then forwards the \emph{computation result}.
%\end{enumerate}

Rather than regarding the traffic of M2O feature as a nuisance, we propose a new physical layer scheme, dubbed STAC (Simultaneous Transmission and Air Computation), to take advantages of the \emph{superposition} nature of EM waves and the M2O transmissions. Our STAC are based on two key observations on the distinguishing features of wireless DCNs.

{\it{Observation 1:}} One feature in DCs is that these M2O sessions are generally followed by certain further computations at the destination nodes. These computations  normally satisfy the \emph{commutative} and \emph{associative} operational laws, with weighted summation being the typical case (e.g., in linear network coded storage \cite{Dimakis_NCstorage_2010} and MapReduce-based machine learning  \cite{chu_mapreduce_learning:2006} applications). This opens up the possibility of dividing a whole computation task into several sub-tasks that can be conducted at the intermediate relay nodes, rather than demanding the final destination do all the jobs.  In other words, instead of forwarding all the received digits, the relay could perform some intermediate computation and then forward only the output of the computation, thereby utilizing the bandwidth more efficiently {\footnote{This can be regarded as a simple extension of the combiner operation from the source node to the relay nodes.}}. Considering that the bottleneck of the development of DCs lies in the DCN, not the compute capabilities of the works nodes, we believe that such Compute-and-Forward (CAF) relay strategy is preferable to the traditional SAF strategy for DCNs.

{\it{Observation 2:}} Another feature in DCs is that the static closed environment, where all the work nodes are closely placed in one relatively small rooms. As a result, the transceiver positions and the channel between them are time invariant. Moreover, with the indirect ceiling-reflection and the 60GHz techniques \cite{Zhou:2012:MMC:2342356.2342440}, the channel between transceivers are indirect Line of Sight (LoS) channel without multi-path effect. These two facts help to easy the cooperative transmissions among the nodes.

With respect to {\it{Observation 1}}, it suffices to illustrate STAC for a particular multiple-access unit as depicted in Fig. \ref{fig_illustrating_model}. Suppose that the receive node $0$ is only interested in the weighted summation $s_0$ of the $K$ source digits
%\footnote{The digit here is an application layer concept, and should be distinguished from the physical layer symbol/signal. The detailed mapping mechanism between them and signal detection method will be discussed in Section 4.}
$s_1,s_2,\ldots,s_K$,
\begin{align} \label{eq:trancomp_math_model}
s_0 = \sum_{i=1}^{K}w_i s_i,
\end{align}
where $w_1, w_2,\ldots,w_K$ are the weight coefficients, and all the quantities here are assumed to be real integers throughout this paper. In STAC,  the $K$ source nodes transmit their digits in the same time slot with appropriately chosen transmit powers, frequencies, phases and times, such that their information bearing EM waves arrive at node 0 in a desired \textit{superimposed} form that can be transformed to $s_0$ directly. As will be shown, this new STAC scheme significantly improves  the separate P2P/SAF-based-transmission and computation strategy, in terms of bandwidth and energy efficiencies. Additionally, in the general case when node 0 needs to fully recover the original $K$ source digits, e.g., for performing some computation other than weighted summation, one can still apply STAC by properly designing a set of \emph{pseudo} coefficients $\{w_1, w_2,\ldots,w_K\}$ such that the original digits $s_1,s_2,\ldots,s_K$ can be extracted from the received $s_0$.

To enable STAC, accurate channel state information (CSI) and perfect frequency/time synchronization among the transceivers are needed, both of which may be difficult to obtain in general wireless networks. Thanks to {\it{Observation 2}}, however, the CSI in a DC is nearly time-invariant and can be accurately estimated.

To accomplish the synchronization, as another contribution, this paper novelly proposes to use wired connections among all the work nodes to provide the wireless transceivers external reference signals (e.g., a high quality external clock signal) \cite{argos_li_2012}, based on an enhanced Software Defined Network (SDN) architecture \cite{McKeown_openflow_2008}.
It should be pointed out that, the wired connections here are distinguished from the information transmission links in a wired DCN. The former are dedicated and solely responsible for control signals, not requiring the high bandwidth and random traffics as in the latter, and thus will not cause the aforementioned problems encountered by wired DCNs. We also remark that to build up such a wired control network in DCs is plausible considering that the work nodes are usually compactly piled up in a dedicated room of limited size. As a by-product, it will also reduce the DCN operation cost by eliminating the need of using individual oscillators at the transceivers.

\section{Motivating Examples}
%This section details some DC application examples mentioned in Section 1 that motivate our STAC scheme.

Two major DC applications are i) distributed file storage, e.g., GFS \cite{Ghemawat_GFS_2003} and Hadoop Distributed File System (HDFS) \cite{hdfs_2010}, and ii) parallel big data processing, typically based on the MapReduce style framework \cite{Dean_Mapreduce_2008}. We now present three detailed DC application examples mentioned in Section 1 that motivate our STAC scheme, where the first two correspond to GFS and MapReduce, respectively, and the last one shows the flexibility of STAC for general applications. Again, with task division, we can concentrate our discussions on the multiple-access unit depicted in Fig. \ref{fig_illustrating_model}.

%Notably, with the task division approach clarified in Section 1, in the following we only focus on the operation of a single multiple-access unit without loss of generality.

%A common feature of these two applications lies in that the desired processing function is normally communicative and associative (e.g., recall that a typical processing task in DC is in the form of weighted summation).
%This feature then opens up the possibility of splitting a whole task into several sub-tasks that are conducted at intermediate relay nodes, rather than demanding the final destination doing all the jobs. With such an option,
%
%
% within different multiple-access units

\textbf{Network Coded Storage.} Due to the nonnegligible node failures in a DC \cite{Ghemawat_GFS_2003}, in distributed storage systems, a big file is usually divided into many fixed-length data blocks that are further protected by multiple replicas stored at different work nodes.

For storage efficiency, network code (or erasure code) can be applied \cite{Dimakis_NCstorage_2010, wu:nc_storage:5402495, Papailiopoulos:2012:6195703}, where each node stores the network coded data blocks rather than their original forms.  When a data block is lost due to the node failure, it can be reconstructed at a new node by performing the following algorithm digit-by-digit:
\begin{algorithm}[H]
\caption{Network Coded Recovery} \label{eq:network_coding_recover}
\begin{algorithmic} [1]
   \State $s_0 = \sum_{i=1}^{K}w_is_i $
    \State $s_0 \gets s_0 \bmod 2^q$
\end{algorithmic}
\end{algorithm}
\noindent where $s_0$ denotes a digit from the lost data block requiring recovery,  $s_1,s_2,\ldots,s_K$ are digits from the data blocks stored at the other nodes, $w_1, w_2,\ldots,w_K$ are the network coding coefficients, and the modulo operation is due to the finite field size $2^q$. Clearly, with STAC, we can achieve Step 1 of the algorithm directly.

%Suppose nodes $0,1,\ldots,K$ are as depicted in the multiple-access unit of Fig \ref{fig_illustrating_model}.
%Instead of letting the $K$ nodes successively transmit $\{s_i, i=1,2,\ldots,K\}$ to node 0 for further computations as in the above algorithm, with STAC in \eqref{eq:trancomp_math_model}, we can achieve this directly.

%\begin{algorithm}
%\caption{Euclid's algorithm}\label{euclid}
%\begin{algorithmic}[1]
%%\Procedure{Euclid}{$a,b$}\Comment{The g.c.d. of a and b}
%    \State $r\gets a\bmod b$
%   \If{sfdfdsa}
%   \State sdfsadf
%   \EndIf
%   \While{$r\not=0$}\Comment{We have the answer if r is 0}
%      \State $a\gets b$
%      \State $b\gets r$
%      \State $r\gets a\bmod b$
%   \EndWhile\label{euclidendwhile}
%  % \State \textbf{return} $b$\Comment{The gcd is b}
%%\EndProcedure
%\end{algorithmic}
%%\end{algorithm}

\textbf{MapReduce Based Data Processing.} Popularized by Google, MapReduce is a dominant parallel big data processing tool in DCs. In MapReduce model, when the map nodes finish the processing, their outputs with the same key will be  sent to a specified reduce node for the final computations. Such computations are also typically in the form of weighted summations \cite{ranger_mapreduce_on_multicore_2007, Dean_Mapreduce_2008}, e.g., for all machine leaning algorithms fitting the statistical query model \cite{chu_mapreduce_learning:2006}, scientific processes \cite{seo_matrix_mapreduce_2010, Liu_matrix_factorization_2010}, parallel $K$-means\cite{zhao_parallel_kmeas_2009}, prefix sum and brute-force sorting \cite{goodrich_sorting_2011}, documents similarity comparisons \cite{Elsayed_document_compare_2008}, etc. Again, our STAC scheme can be applied to achieve the simultaneous transmissions and computations efficiently.

\textbf{General Case.} In DCs, there are quite a few other applications, in which the additional task division does not applied. In such cases the receive node 0 needs the original source digits, one can appropriately design a set of pseudo coefficients $\{w_1,w_2,\ldots,w_K\}$ such that the source digits $s_1, s_2, \ldots, s_K$ can be extracted from $s_0$. In particular, suppose for each $i=1,2,\ldots,K,$
% the source digit $s_i, i=1,2,\ldots,K$ satisfies
 $0 \leq s_i \leq 2^q-1,$  then choosing $w_i=2^{q(i-1)}$ yields
\begin{align*} s_0 = \sum_{i=1}^{K} 2^{q(i-1)}s_i, \end{align*}
based on which all the source digits can be extracted with the following algorithm:
\begin{algorithm}
\caption{Source Digits Extraction}\label{alg:individual_symbol}
\begin{algorithmic}[1]
\State $i \gets 1$
\While{$i \leq K$}
    \State $s_i \gets s_0 \mod \ 2^q  $
    \State $ s_0 \gets (s_0 -s_i) / 2^q  $
    \State $i \gets i+1$
\EndWhile
\end{algorithmic}
\end{algorithm}

\section{System Framework with STAC}

%, where multiple work nodes transmit to one receive node simultaneously, including channel model, system architecture, mapping between information digits and modulated symbols, signal detection and energy efficiency issues.

\subsection{A Basic STAC Unit}
\begin{figure}
\begin{center}
\includegraphics[width=3.0in]{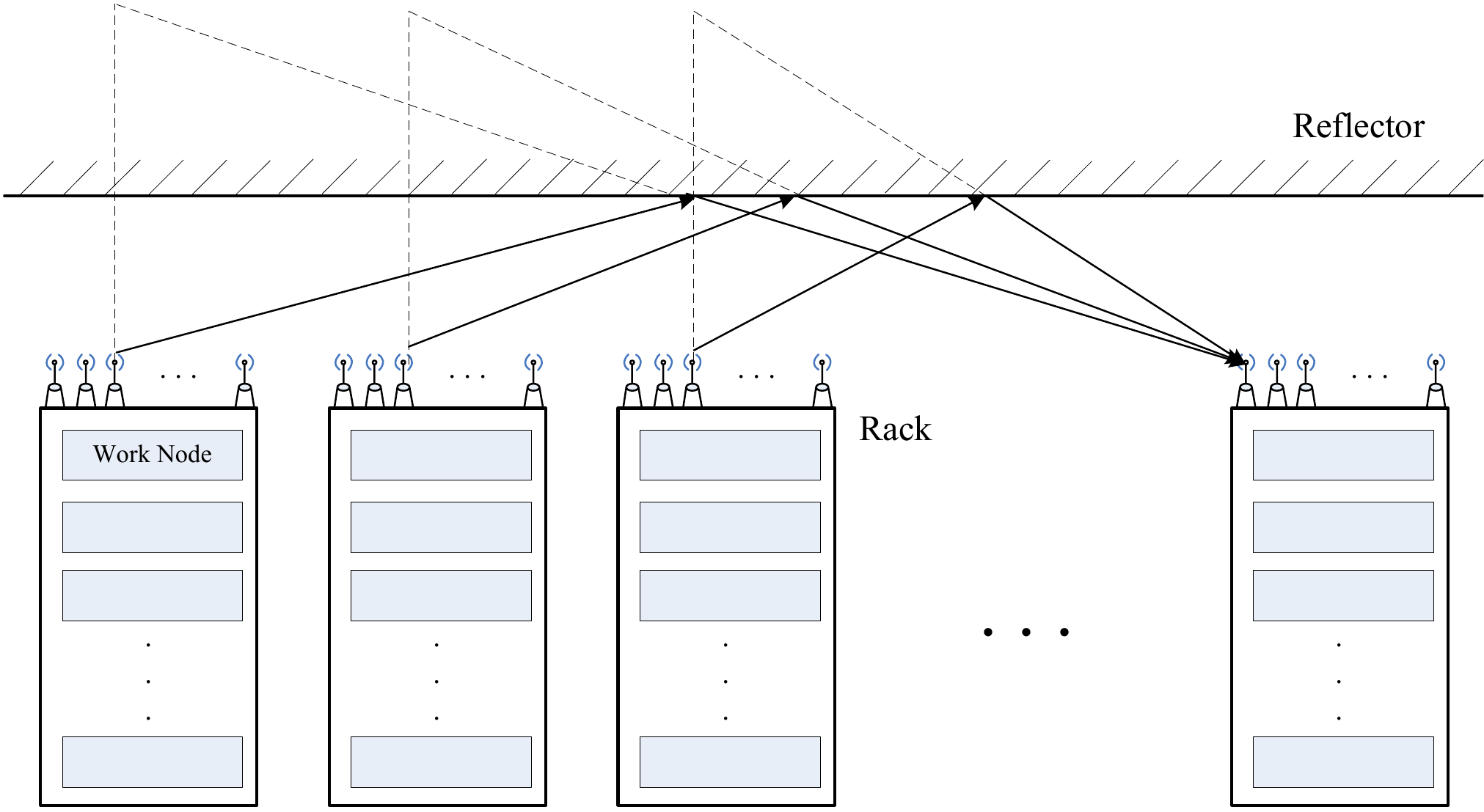}
\caption{A typical wireless DC layout.}
\label{fig_communication_model}
\end{center}
\end{figure}

STAC is a general physical layer scheme that can be applied to wireless DCs with any structure, carrier frequency, etc. For illustration, consider a typical layout of the wireless DC as shown in Fig. \ref{fig_communication_model}, where each rack contains multiple work nodes and has an antenna array mounted on its top to communicate with other racks (communications within a rack are accomplished with intra-rack connections) \cite{Shin:2013:FCW:2578911.2578935}. As in \cite{Zhou:2012:MMC:2342356.2342440}, ceiling-reflecting and 3D beamforming techniques are adopted to achieve an indirect LoS link between any two antenna arrays without causing interference to others.

Suppose $K$ work nodes (in $K$ different racks) need to transmit their digits $s_1, s_2, \ldots, s_K$ to node 0 for computing the weighted summation as in \eqref{eq:trancomp_math_model}. The operating principle of STAC is illustrated in the following.

Each source node $i$ maps its digit $s_i$ to a baseband modulated complex symbol $d_i$, and then up converts the symbol $d_i$ to a passband signal given by
$$\sqrt{P_i} e^{-j\theta_i} d_i(t) e^{-jf_ct},$$
where $\theta_i$ and $\sqrt{P_i}$ are the pre-equalizing phase and amplitude coefficients, respectively. Suppose each node $i$ transmits at time $t_i$ using 3D beamforming, then the received passband signal $ {y}(t)$ can be expressed as
%steers its antenna array to the receiver and transmits at time $t_i$, then the received passband signal can be expressed as
\begin{align*}
\sum_{i=1}^{K} h_ie^{j\theta'_i} \sqrt{P_i} e^{-j\theta_i} d_i(t-t_i-\tau_i) e^{-jf_c(t-t_i-\tau_i)}
+ n(t)
\end{align*}
where $h_ie^{j\theta'_i}$ is the equivalent complex channel coefficient from node $i$ to $0$, $\tau_i$ is the propagation delay for node $i$, and $n(t)$ is a Gaussian noise of variance $\sigma^2$ for both the real and imaginary dimensions. With accurate CSI, one can set
\begin{align}\theta'_i=\theta_i \text{ and } t_i = t_0-\tau_i,\label{eq:set1}
\end{align}
such that
the received signal simplifies to
\begin{align*}
%\label{eq_passband_model}
 {y}(t) = \sum_{i=1}^{K} h_i \sqrt{P_i}   d_i(t-t_0) e^{-jf_c(t-t_0)} + n(t),
\end{align*}
which, after down conversion and sampling at time $t=t_0$, yields the baseband symbol\footnote{The $h_i$ in \eqref{eq_baseband_model} are real variables, so that the real and imaginary parts of symbol $y$ can be separated. The sequel of this paper will only consider the real part for simplicity.}
\begin{align} \label{eq_baseband_model}
y = \sum_{i=1}^{K} h_i \sqrt{P_i} d_i  + n.
\end{align}
Clearly, if each node $i$ sets
\begin{align}P_i=(w_i/h_i)^2,\label{eq:set2}\end{align}
then after eliminating the noise, node 0 can construct the desired digit $s_0$ as in \eqref{eq:trancomp_math_model} from the symbol $y$ in \eqref{eq_baseband_model}.

With the above described principle, we can find that the time/frequeny synchronization and pre-equalization, such as \eqref{eq:set1} and \eqref{eq:set2},  are essential for our STAC. They can be realized based on an enhanced SDN architecture as shown in the next.

%
%\section{Technique Design Issues}

\subsection{An Enhanced SDN Architecture}
%As shown in above subsection, there are a lot of things needs to be synchronized and pre-equalized, such as carrier frequency, timing, channel equalization and steering configuration. Considering current DC computation architecture,  we propose a enhanced software defined architecture \cite{Zhou:2012:MMC:2342356.2342440} to ease the synchronization and pre-equalization jobs, as shown in Figure \ref{fig_system_architecture}.

\begin{figure}[tbp]
\begin{center}
\includegraphics[width=3.0in]{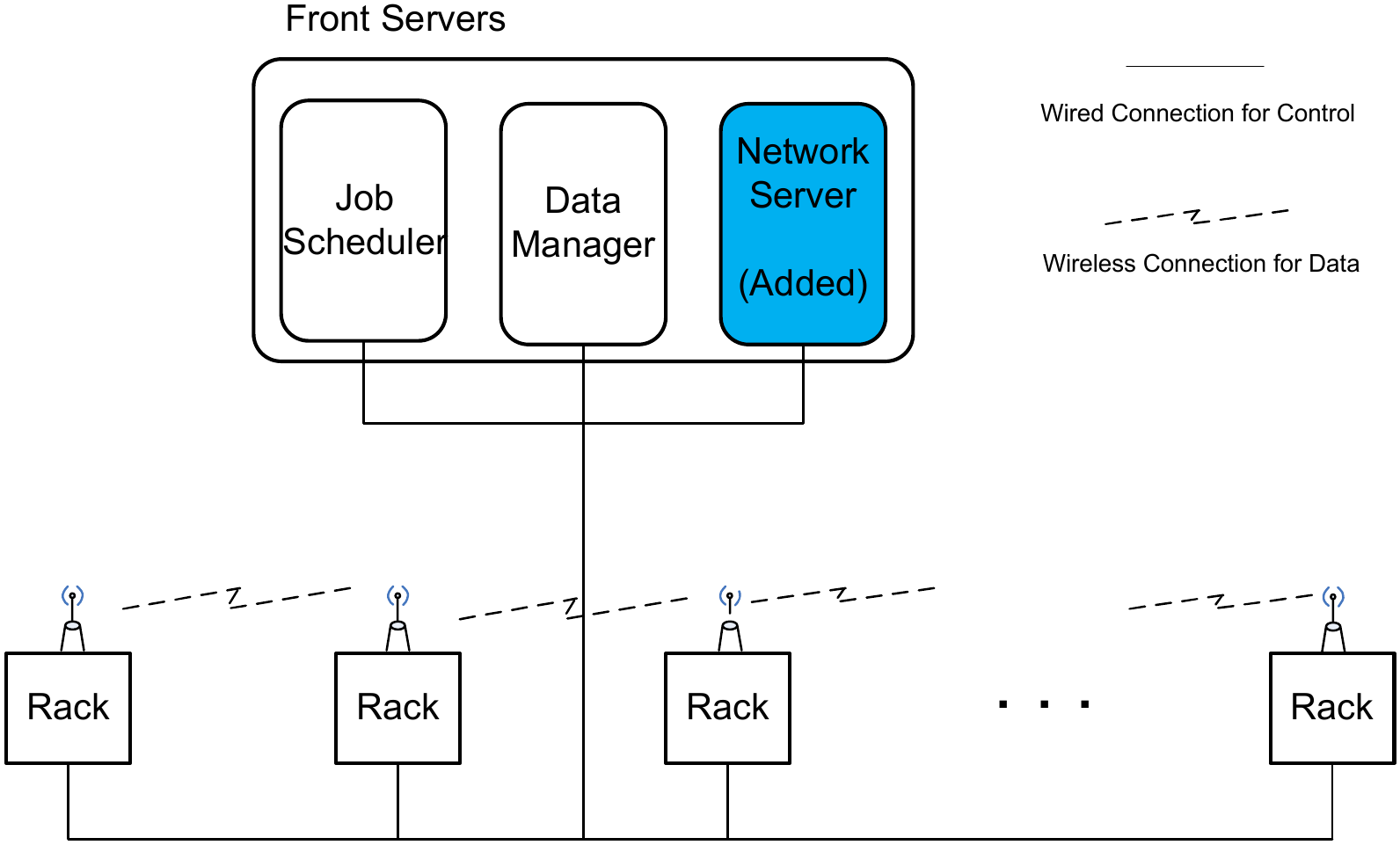}
\caption{An enhanced SDN architecture.}
\label{fig_system_architecture}
\end{center}
\end{figure}

%\textbf{Separated Control Channel:}
The DC generally works in a  centralized control manner, where the front servers, including the job scheduler and data manager, manage all the work nodes. In current DCNs, control signals and data traffic share the same network. Here, we propose to use a dedicated low bandwidth wired control network with an added network server as shown in Fig. \ref{fig_system_architecture}, based on an enhanced SDN architecture.
As mentioned in Section 1, the feasibility of establishing the wired control network is endorsed by the limited DC size and the fixed node locations.

Our SDN architecture is an enhanced one in the sense that, it not only accomplishes networking control as in general SDNs, but also also provides the wireless transceivers the physical and upper layer configurations to enable STAC, including the synchronization information, the physical layer parameters such as powers, frequencies, phases and times, and the scheduling/routing information.

%We establish wired connections among all the nodes to convey the control information to administrate the overall wireless network. One network server (controller in SDN) connects to all the nodes through a one-layer switch. Then, this network server controls the wireless transmission of all the work nodes through this wired control channel. On the other hand, the wireless network is dedicated to the traffic information transmission in the DC.

\textbf{Synchronization with External Reference Signals.}
External reference signals are provided to all the transceivers for synchronization. These include a high quality external clock signal, with which individual crystal oscillators at the transceivers are no longer needed and the operation cost can be thereby reduced. These reference signals can also help calibrate the wireless transceivers, e.g., reduce the errors induced from the device hardware differences \cite{argos_li_2012}.

\textbf{Physical Layer Parameters.}
The network server maintains a connection information table that stores important physical layer parameters for each connection, such as the transmission delay $\tau$, channel coefficient $he^{-j\theta}$ and the steering vectors required for 3D beamforming.
%Such connection information will be updated periodically or upon request.
When a  transceiver fails (or a new one comes in), it informs the network server through the control network to remove (add) it from (to) the connection information table.

\textbf{Scheduling/Routing.}
Also maintained by the network server is a table storing the scheduling/routing information.
When a current task finishes or a new one needs to start, the job scheduler informs the network server to update the scheduling/routing information table, and then the network server will do the corresponding coordinations among all the work nodes involved.

\section{Physical Layer Issues}

\subsection{Modulation-Demodulation Mapping}
The modulation for STAC is the same as that for P2P channels. However, their demodulation mappings are subtly different: STAC demodulation maps a superimposed symbol, which may even not belong to the transmit symbol sets, to the summation of the digits, whereas the P2P channel demodulation maps a particular symbol from the transmit symbol set to the corresponding digit.

\textbf{STAC Modulation.} Specifically, writing node $i$'s digit $s_i$ into the bit sequence form yields
$$[ s_i(1),s_i(2),\ldots, s_i(l),\ldots, s_i(L) ]$$
where $s_i(l)$ is the $l$-th bit,  $L$ is the sequence length, and
\begin{align*}
s_i = \sum_{l=0}^{L-1} {2^{l}s_i(l)}.
\end{align*}
For modulation, assume BPSK (Binary Phase Shift Keying)\footnote{STAC also applies with other modulations such as QPSK, QAM, OOK, OFDM, etc. This paper only considers the simplest BPSK due to the same reason mentioned in Footnote 1.} without error correction coding throughout this paper.
% throughout this paper.
%Each dimension of QPSK, real or the imaginary, is equivalent to a BPSK (Binary Phase Shift Keying) symbol that can represent a bit $s_i(l)$, and one only needs to consider the real dimension due to the reason mentioned in Section 3.1.
At node $i$, each bit $s_i(l)$ is modulated to a symbol $d_i(l) \in \{-1,+1\}$ as
$d_i(l)=1-2\times s_i(l).$

\textbf{STAC Demodulation.}
After the $l$-th transmission and the removal of noise with signal detection, the received superimposed symbol can be written as
\begin{align} \label{eq:noise_less_symbol}
y(l)=\sum_{i=1}^{K}h_i\sqrt{P_i}d_i(l)\end{align}
By setting the transmit power\footnote{With the unit power of $d_i$ in BPSK, the transmit power $P_i|d_i|^2$ simply equals $P_i$.}
$P_i = (w_i/h_i)^2$, one has
\begin{align}
y(l)=\sum_{i=1}^{K}w_i d_i(l), \label{eq:symbol}\end{align}
which, through the operation
$$  \frac{1}{2}\left(\sum_{i=1}^{K}w_i -y(l)\right) ,$$
yields the summation $\sum_{i=1}^{K}w_i s_i(l).$
Finally, the desired digit can be constructed as
\begin{align*}
%\label{eq_transum_modul_nonoise}
\sum_{l=0}^{L-1}2^l \sum_{i=1}^{K}w_i s_i(l)&= \sum_{i=1}^{K} w_i \sum_{l=0}^{L-1}2^l  s_i(l)=  \sum_{i=1}^{K}w_i s_i.
\end{align*}

\subsection{Signal Detection}
We now present a simple signal detection scheme for removing the noise in \eqref{eq_baseband_model} to obtain \eqref{eq:noise_less_symbol}, and analyze its corresponding SER (Symbol Error Rate). It suffices to consider only one of the $L$ transmissions, and hence the index $l$ as in the last subsection will be omitted.

Specifically, view the symbol $\sum_{i=1}^{K}w_id_i$ in \eqref{eq:symbol}  as a point of a non-standard PAM (Pulse Amplitude Modulation) constellation that results from the weighted superposition of the transmit BPSK constellations and hence may have unequal distance between different adjacent constellation points. A simple detection scheme is to quantize the $y$ in \eqref{eq_baseband_model} to its nearest constellation point.  Let $\pi$ be a permutation on $\{1,2,\ldots,K\}$ such that $w_{\pi(j_1)}\leq w_{\pi(j_2)}$, $\forall j_1\leq j_2$. We have the following theorem regarding the SER with such detection.
\newtheorem{theorem}{Theorem}
\begin{theorem}\label{Thm:SER_stac}
The SER with the nearest point detection is upper bounded by
\begin{align}\mbox{SER}_{\text{\tiny{STAC}}} \leq (1-1/2^{K}){\it{erfc}}(1/\sqrt{2}\sigma)\label{E:achieveseq}\end{align}
where ${\it{erfc}}(x)=\frac{2}{\sqrt{\pi}}\int_x^\infty e^{-t^2}\,dt$ is the complementary error function,
$\sigma^2$ is the variance of the noise, and the equality
in \eqref{E:achieveseq}
holds when the distance between any two adjacent constellation points is equal to 2, e.g., when $w_{\pi(j)} = 2^{j-1} \ or  \ 1, \forall j=1,\ldots,K.$
%\begin{align}w_{\pi(j)} = 2j, \forall j=1,\ldots,K. \label{eq:constraint}\end{align}
\end{theorem}
\textbf{Proof Sketch:}
Since $w_i$ are all real integers, the largest SER is attained when the distance between any two adjacent constellation points is 2, which includes the case of $w_{\pi(j)} = 2^{j-1} \ or  \ 1, \forall j=1,\ldots,K.$

%
%Note that the nearest point detection is in general \emph{suboptimal} due to the possible different weight coefficients and the SER in \eqref{E:achieveseq} serves an upper bound. Nevertheless, in the next we will show that even with this suboptimal detection, STAC outperforms the separate P2P-based transmission and computation strategy.

\subsection{Performance of STAC}

The performance of STAC is a tradeoff among SER, energy efficiency and bandwidth efficiency, and is clearly dependent of the weight coefficients. The  \emph{air computation} essence of STAC and its advantage over the separate strategy can be best illustrated in the ideal case of $w_1=w_2=\cdots=w_K=1$, where we will show that for fixed energy efficiency, STAC achieves better SER and significantly improved bandwidth efficiency.
%
% than the separated strategy.

On the other hand, to show that STAC uniformly outperforms the separate strategy, we will consider the pseudo coefficients case as mentioned in Section 2, i.e., $w_{\pi(j)}=2^{j-1},\forall j$. The argument here is that by applying STAC with the pseudo coefficients, one can recover the original $K$ source digits, based on which  summation with any weight coefficients can be computed.
We will show that in this case, STAC achieves better energy efficiency for fixed SER and bandwidth efficiency.

\subsubsection{The Ideal Case}
Suppose $w_1=w_2=\cdots=w_K=1$, which is the ideal case for STAC. The SER of STAC is given in Theorem \ref{Thm:SER_stac}, i.e.,
$$\mbox{SER}_{\text{\tiny{STAC}}} = (1-1/2^{K}){\it{erfc}}(\frac{1}{\sqrt{2}\sigma}).$$
Note that in this case, the resultant receive PAM constellation has only $K+1$, instead of $2^K$, points,
%i.e., $\{-K,  \ldots,-1,0,1,\ldots, K   \}$,
where the decrease of the constellation size is due to the ``air computation''. Or equivalently, viewed from the energy perspective, this advantage is reflected by the fact that the needed transmit power now attains the minimum $P_i=1/h_i^2$ for each node $i$.

For the separate strategy, assume each node $i$ transmits with the same power $P_i=1/h_i^2$ as in STAC. The SER for node $i$  is a standard result, given by $\frac{1}{2} {\mbox{erfc}}(\frac{1}{\sqrt{2}\sigma})$. Combining all the detected $K$ symbols,  the receiver computes $\sum_{i=1}^{K}w_id_i$, and the resultant $\mbox{SER}_{\text{\tiny{SEP}}}$ is characterized in the following theorem.

\begin{theorem}\label{Thm:SER_sep}
The SER with the separate strategy is given lower bounded by
$$\mbox{SER}_{\text{\tiny{SEP}}}\geq \frac{1}{2}-\frac{1}{2}\left(1-{\mbox{erfc}}(\frac{1}{\sqrt{2}\sigma})\right)^K.$$
\end{theorem}
where the equality achieves when $w_1=w_2=\cdots=w_K=1$.

\textbf{Proof Sketch:} The theorem can be proved by noting that the number of erroneous symbols is a binomial random variable with parameters $(K,p)$, and the computation result is wrong if and only if there are odd number of erroneous symbols when $w_1=w_2=\cdots=w_K=1$.

\begin{theorem}
$\mbox{SER}_{\text{\tiny{SEP}}}>\mbox{SER}_{\text{\tiny{STAC}}}$ for any $K\geq 2$.
\end{theorem}

\textbf{Proof Sketch:} Use mathematical induction.

%$$\mbox{SER}_{\text{\tiny{SEP}}}  = \sum_{i=0}^{\lfloor K/2 \rfloor} {{K}\choose {2i+1}} \left( \frac{1}{2} {\mbox{erfc}}(1/\sqrt{2}\sigma) \right)^{2i+1}.$$

%
%
%
%
%\vspace{2mm}
%Now consider the SER with separated transmission and computation. For comparison, assume the same individual transmit power $P_i=(w_i/h_i)^2$ as in STAC.
%Then, the SER for each transmit node is a standard result, given by
%  $$ \frac{1}{2} {\mbox{erfc}}(1/\sqrt{2}\sigma).$$

%\begin{figure}
%\begin{center}
%\includegraphics[width=3.5in]{SER}
%\caption{SER comparison ($K=4$)}
%\label{fig_ser}
%\end{center}
%\end{figure}

%\vspace{2mm}
Therefore, STAC achieves a better SER and simultaneously improves the bandwidth efficiency by a factor of $K$.
Especially, note that as $K\to \infty$, $\mbox{SER}_{\text{\tiny{STAC}}}\to {\mbox{erfc}}(1/\sqrt{2}\sigma)$  whereas $\mbox{SER}_{\text{\tiny{SEP}}}\to 1/2$.
%Fig. \ref{fig_ser} plots $\mbox{SER}_{\text{\tiny{STAC}}}$ and $\mbox{SER}_{\text{\tiny{SEP}}}$ for $K=4$.

To achieve the same bandwidth efficiency, suppose each node transmits $K$ bits in one symbol for separated transmission. Then each node need to increase its transmit power at least by a factor $2^K$, resulting an SER more than $\mbox{SER}_{\text{\tiny{SEP}}}$. In other words, STAC can improve the energy efficiency by a factor more than $2^K$ in
the ideal case.

%Therefore, STAC achieves better SER with the bandwidth efficiency improved by a factor of $K$. Especially note that as $K\to \infty$, $\mbox{SER}_{\text{\tiny{STAC}}}\to 2p$ with $p$ generally small in the high receive SNR (signal-to-noise ratio) regime, whereas $\mbox{SER}_{\text{\tiny{SEP}}}\to 1/2$ regardless! Fig. \ref{fig_ser} plots $\mbox{SER}_{\text{\tiny{STAC}}}$ and $\mbox{SER}_{\text{\tiny{SEP}}}$ for $K=4$.

%
%As we can see, STAC achieves better SER with the bandwidth efficiency improved by a factor of $K$.

%\subsection{Refined STAC}
%%Previous subsections focus on the feasibility and bandwidth efficiency of STAC with given coefficients  $w_i$.
%This part discusses the energy efficiency of STAC, including the general case  where individual digits $s_i$ are needed at the receiver and the tradeoff between the energy efficiency and bandwidth efficiency.

\subsubsection{Pseudo Coefficients Case}
Consider a set of pseudo coefficients $w_{\pi(j)}=2^{j-1},\forall j.$ To minimize the total transmit power $\sum_{i=1}^K (w_i/h_i)^2$ with STAC, we allocate these coefficients among the $K$ nodes such that $h_{\pi(j_1)}\geq h_{\pi(j_2)}$, $\forall j_1\leq j_2$. Assuming STAC is completed within unit time, the total transmit energy $E_{\tiny{STAC}}$ is given by
%$w_i$ is determined by the application and can be any values. We now discuss the worst case where the weight coefficients are $1, 2, 4, \ldots, 2^{K-1}$. We call it worst case since individual digit $s_i$ can be obtained from such weight coefficients to make any desired computation. We also show that STAC with such worst weight coefficients also outperforms the separated transmission and computation scheme.
%
%We first calculate the needed energy for STAC transmitting one bit to the receiver within unit time.
%
%In the general case, the receiver detects each digit $s_i$ and can processes them with any needed manipulation.
%Here, the coefficients for each $w^*_i$ needs to be appropriately chosen. First, the coefficients set should satisfy the constraint as in \eqref{eq:constraint}, which is $1, 2, 4, \ldots, 2^{K-1}$. Then, to minimize the total transmit power $\sum_{i=1}^K (w^*_i/h_i)^2$, we need to associate the largest $w^*_i$ to the node with largest channel coefficient $h_i$.
%
%Finally, the total transmit energy for the general case is
\begin{align}
E_{\text{\tiny{STAC}}}=\sum_{j=1}^K \big( 2^{(j-1)}/h_{\pi(j)}\big)^2.
\end{align}
%where, similar to $\pi(i)$, $\pi'(i)$ is another permutation function such that $h_{\pi'(j_1)} \leq h_{\pi'(j_2)}, \forall j_1 \leq j_2$.

%For any system with given $w_i$, if the total power $\sum_{i=1}^K (w_i/h_i)^2 \geq P^*$, we should use $w^*(i)$, instead of $w_i$, for STAC transmission, and the receiver can recalculate the desired $\sum_{i=1}^K w_is_i$ after obtaining each $s_i$.
%In summary, STAC can finish the transmission with at most a total power of $P^*$,.

We now calculate the total energy needed $E_{\text{\tiny{SEP}}}$ for the separate strategy assuming that each node transmits 1 bit to the receiver within $1/K$ time to maintain the same bandwidth efficiency as STAC.
For the separate strategy to achieve the similar SER as STAC, the distance between any adjacent receive constellation points also needs to be 2, in which case node $i$'s transmit power is given by
$$P_i = \sum_{j=1}^K \big( 2^{(j-1)}/h_i \big)^2. $$
Therefore, the total energy needed is
\begin{align} \label{eq:energy_efficiency}
E_{\text{\tiny{SEP}}}&= \frac{1}{K}\sum_{i=1}^K  \sum_{j=1}^K \big( 2^{(j-1)}/h_i \big)^2
%&  =\frac{1}{K}\sum_{i=1}^K 1/h_i^2  \sum_{j=1}^K \big( 2^{(j-1)} \big)^2.
\end{align}
where the factor $1/K$ accounts for the transmission time of each node.

\begin{theorem}\label{Thm:Energy}
$E_{\text{\tiny{SEP}}} \geq E_{\text{\tiny{STAC}}}$, where the equality holds only when $h_i$ are the same for all $i$.
\end{theorem}
\textbf{Proof Sketch:} The proof utilizes the important fact that $w_{\pi(j_1)}\leq w_{\pi(j_2)}$ and $h_{\pi(j_1)}\geq h_{\pi(j_2)}$, $\forall j_1\leq j_2$.

From Theorem \ref{Thm:Energy}, it can be concluded that STAC performs uniformly better than the separate strategy for any set of weight coefficients. This is because even requiring STAC to fully recover the original $K$ source digits leads to better energy efficiency than the separate strategy, for fixed SER and bandwidth efficiency.

\subsubsection{Discussion}
The above analyzes two extreme cases of the weight coefficients. In general, depending on the specific weight coefficients, one has the freedom of dividing the $K$ nodes into $M$ groups ($1\leq M\leq K$), and letting each group transmit using STAC separately, to achieve a tradeoff between the bandwidth efficiency and energy efficiency.

\section{Conclusion}
%
%Wireless data center network is a new kind of wireless networks, with compact space, random and heavy data traffic, static channel conditions and possible . Therefore, wireless DCN needs its own transmission and networking design, different from the traditional cellular or WiFi networks, to discover its full potentials.

The wireless DCN differs from general wireless networks in that it has large amounts of M2O sessions, which are normally followed by further computations at the destinations, with weighted summation being the typical case. Recognizing this, we have proposed a novel physical layer scheme STAC that achieves simultaneous transmissions and computations over the air, and an enhanced SDN architecture to enable it. It is demonstrated that with STAC used, both the bandwidth and energy efficiencies can be significantly improved.

\bibliographystyle{unsrt}
\bibliographystyle{abbrv}
\bibliography{./final_refs}  % sigproc.bib is the name of the Bibliography in this case
\end{document}